# The potential stickiness of pandemic-induced behavior changes in the United States

Deborah Salon[1]\*, Matthew Wigginton Conway[1], Denise Capasso da Silva[2], Rishabh Singh Chauhan[3], Sybil Derrible[3], Kouros Mohammadian[3], Sara Khoeini[2], Nathan Parker[4], Laura Mirtich[1], Ali Shamshiripour[3], Ehsan Rahimi[3], and Ram Pendyala[2]

[1]School of Geographical Sciences and Urban Planning, Arizona State University; Tempe, AZ, USA.

[2]School of Sustainable Engineering and the Built Environment, Arizona State University; Tempe, AZ, USA.

[3]Department of Civil, Materials, and Environmental Engineering, University of Illinois at Chicago; Chicago, IL, USA.

[4]School of Sustainability, Arizona State University; Tempe, AZ, USA.

\*Corresponding author. Email: dsalon@asu.edu

**Email:** dsalon@asu.edu

**Author Contributions:**

Conceptualization: DS, RP, AM, SD, MWC, DCS, RSC, LM, ER, AS
Methodology: DS, MWC, DCS, RSC, RP, AM, SD, LM, SK, ER, AS
Investigation: MWC, DCS, RSC, DS, SD, AM, RP, ER, AS
Visualization: DS
Funding acquisition: RP, DS, SD, AM, MWC
Project administration: DS, AM, SD, RP
Supervision: DS, AM, SD, RP
Writing – original draft: DS, MWC
Writing – review & editing: DS, MWC, RP, NP, SD, DCS, SK, AM, LM, ER, AS, RSC

## Abstract

Human behavior is notoriously difficult to change, but a disruption of the magnitude of the COVID-19 pandemic has the potential to bring about long-term behavioral changes. During the pandemic, people have been forced to experience new ways of interacting, working, learning, shopping, traveling, and eating meals. A critical question going forward is how these experiences have actually changed preferences and habits in ways that might persist after the pandemic ends. Many observers have suggested theories about what the future will bring, but concrete evidence has been lacking. We present evidence on how much U.S. adults expect their own post-pandemic choices to differ from their pre-pandemic lifestyles in the areas of telecommuting, restaurant patronage, air travel, online shopping, transit use, car commuting, uptake of walking and biking, and home location. The analysis is based on a nationally-representative survey dataset collected between July and October 2020. Key findings include that the new normal will feature a doubling of telecommuting, reduced air travel, and improved quality of life for some.

## Introduction

Disruptions in our lives present opportunities to learn and practice new ways of doing things, and to re-evaluate old choices and habits (1). The COVID-19 pandemic has been perhaps the largest disruption event in modern human history. Nearly every human on the planet has been forced to modify their habits to adjust to the pandemic, creating an opportunity for long-term change. Importantly, the pandemic has coincided in time with the widespread availability of technologies such as broadband internet service and videoconferencing, as well as many app-based services available through mobile phones.

To provide insights into the potential stickiness of pandemic-induced behavior changes, we developed an extensive survey and collected 7,613 responses in the United States (U.S.) between July and October 2020 (2). The dataset is weighted to be representative of U.S. adults, and captures pre-pandemic, pandemic-era, and expected future behavior in the areas of telecommuting, restaurant patronage, air travel, online shopping, transit use, car commuting, uptake of walking and biking, and home location.

We compare respondent expectations about their own future choices to their pre-pandemic lifestyles, contributing evidence-based estimates of how much pandemic-era changes may persist in the long run. We focus on expected changes that will be especially consequential for the U.S. economy. Statistical modeling to ascertain the socioeconomic and geographical correlates of these changes is left for future work.

Although we recognize that stated intentions do not always accurately predict future choices, both the survey's design and the choice context itself alleviate this concern. The survey instrument prompted respondents to provide reasons when they reported that they expect to behave differently post-pandemic than was their pre-pandemic norm. These questions served both as a check on whether a change was actually expected and provided information that informs whether the change is likely to stick.

Further, respondents understand the choice context well. They experienced one lifestyle pre-pandemic, their daily lives changed during the pandemic, and our future-looking questions ask them how they plan to mix and match the two ways of life. Respondents have experience with both lifestyles as well as time to reflect on this question during the pandemic, so their answers are well-informed. One of our survey questions provides direct evidence of the "stickiness" of pandemic-induced behavior change; more than 70% of respondents indicated there were aspects of pandemic life they would like to continue.

Survey data documenting differences between pre-pandemic choices and expectations for the post-pandemic future represent the direct, or partial equilibrium, effects of the pandemic. Substantial shifts in choices, however, will cause secondary effects to cascade through the economy, and government policies could shift as well. Estimating these secondary effects is beyond the scope of this Brief Report. We invite others to use this dataset (3) to calibrate general equilibrium models that can provide predictions of both primary and secondary effects.

## Telecommuting and its consequences

The most transformative long-term change identified in our data is a large increase in telecommuting. We asked respondents whether they expect to have the option to telecommute post-pandemic, and if so, how often they expect to do so. Therefore, answers reflect individual preferences tempered by expectations about what their employers will allow. The fraction of workers who expect to telecommute at least a few times each week is double that of the pre-pandemic period, increasing from 13% to 26% (Fig 1).

A shift to telecommuting is important for its direct impacts on quality of life, worker productivity, and commuting. Among those new to telecommuting at least a few times a week during the pandemic, two-thirds identified telecommuting and/or commuting less often as key features of pandemic life they would like to continue into the future. More than 70% of those new to regular telecommuting report that their productivity has stayed the same or improved during the pandemic, consistent with pre-pandemic research (4). This is remarkable, since many pandemic-era telecommuters are juggling childcare and have suboptimal working environments.

The long-term increase in telecommuting is not equitably distributed across the population. Among workers who were not frequent telecommuters pre-pandemic, those who hold a bachelor's degree or live in households earning over $100,000 per year are twice as likely to expect to telecommute at least a few times a week post-pandemic. Thus, these quality of life improvements will flow primarily to high-income, highly-educated individuals.

The direct impacts of telecommuting on car commuting are substantial. We estimate that less frequent commuting (Fig 1) will reduce car commute kilometers by approximately 15%. The fraction of commuters who choose the car as their primary commute mode is not expected to change substantially.

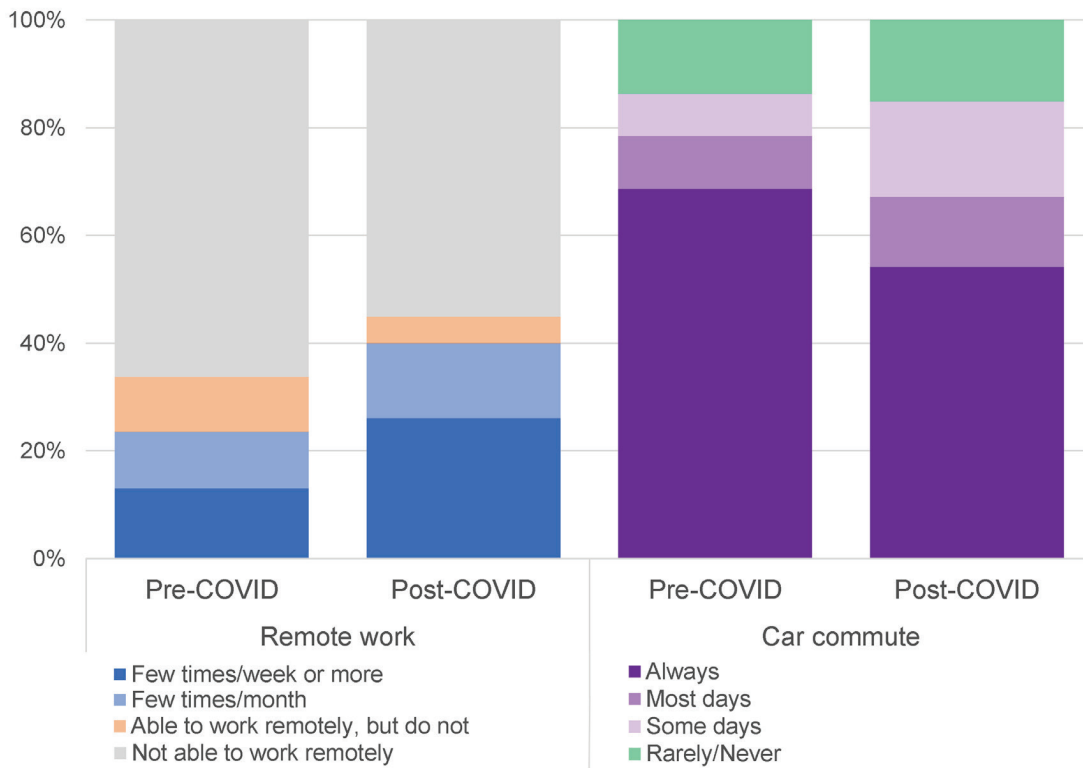

**Figure 1. Key remote work shifts.** Sample sizes: remote work = 4,554 employed adults; car commute = 3,217 commuters.

Telecommuting also impacts transit demand. Though transit systems carried just 5% of U.S. commuters (5), commuting accounted for about half of all transit trips pre-pandemic (6). Our data suggest nearly a 40% decline in transit commute trips post-pandemic, relative to pre-pandemic. Of this decline, about half can be attributed to changes in commuting frequency, 40% comes from a net shift among transit commuters toward the private car, and the remaining 10% comes from shifts to other modes.

A shift to telecommuting will have indirect effects on many aspects of our economy. There is likely to be reduced demand for office space and downtown parking. Patronage of office-district businesses is likely to decrease. Restaurants will continue to be hard-hit. Our data suggest that the number of people who plan to dine in restaurants at least a few times each week will decrease by more than 20% post-pandemic, compared to the pre-pandemic era. Since the restaurant industry employed 8% of U.S. workers pre-pandemic (7, 8), a decrease in restaurant patronage translates to a significant economic hardship for service workers.

**A paradigm shift in air travel**

Air travel demand dropped 95% at the height of the pandemic, and has only rebounded to 38% of its pre-pandemic level as of February 2021 (7). Our data indicate that more than 40% of business travelers expect to travel less frequently post-pandemic (Fig 2). Of those reducing business travel, two-thirds attribute this change to new realizations that are likely to stick, primarily about the utility of videoconferencing. Personal air travelers also expect to fly less (Fig 2), but nearly half of these reductions are caused by pandemic-related concerns that will likely soon fade.

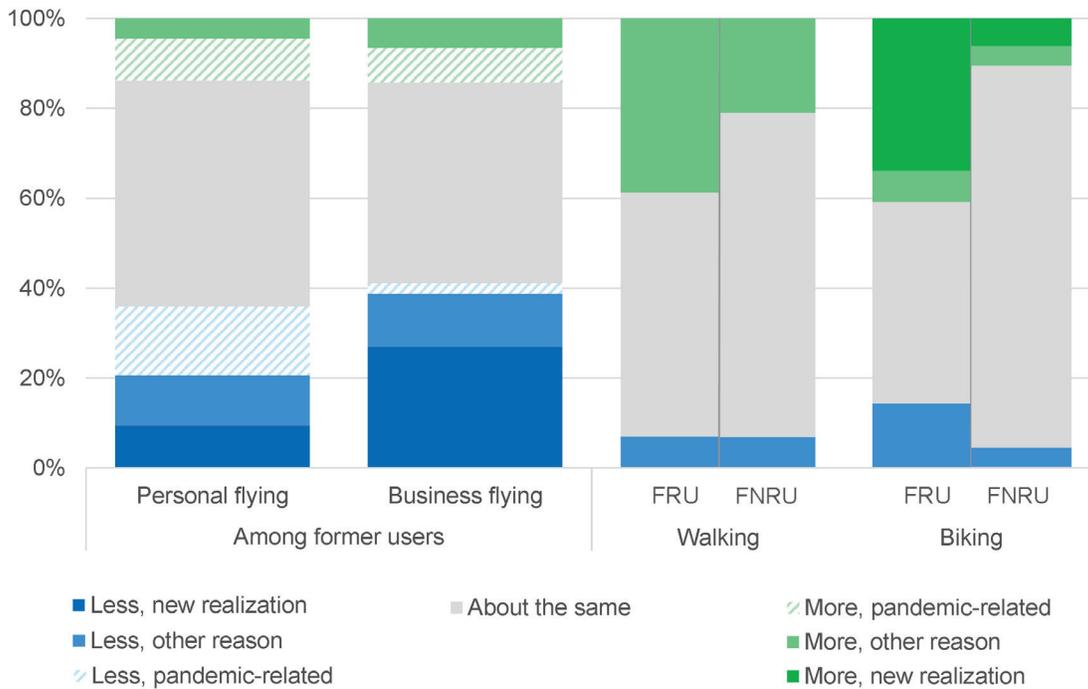

**Figure 2. Pre- to post-pandemic expected shifts in flying, walking, and biking, with reasons.** A sizable fraction of survey respondents selected pandemic-related reasons for their expectations about their future choices. Full details available in the supplemental materials. FRU=Former Regular Users, FNRU=Former Non-Regular Users. Sample sizes: personal flying=5,313 former flyers; business flying=1,676 former flyers; walking, FRU=3,750; walking, FNRU=3,794; biking, FRU=1,000; biking, FNRU=6,395.

**Accelerated growth of online shopping for groceries**

The pandemic has accelerated the uptake of online grocery shopping, nearly doubling the fraction of grocery spending done online (10). We analyzed survey responses from those who tried online grocery shopping for the first time during the pandemic. Approximately half expect to continue to grocery shop online at least a few times a month post-pandemic, but nearly 90% of them also expect to shop in-store for groceries at least a few times a month. This suggests that online grocery shopping does not completely replace in-store shopping, although it may reduce its frequency. Among all U.S. residents, 30% expect to grocery shop online at least a few times a month post-pandemic, up from 21% pre-pandemic.

Our data show online shopping for durable goods following a pre-existing upward trend (11). 63% expect to shop for durable goods online at least a few times a month post-pandemic, compared to 59% before the pandemic.

**Marked increases in walking and bicycling**

Biking and walking have increased during the pandemic in many U.S. cities (12), a change that improves both transport sustainability and public health. Post-pandemic, 30% of U.S. residents plan to take walks more frequently than they did before the pandemic, and nearly 15% plan to bike more (Fig 2). These results include walking and biking for both transportation and recreation, with those who were frequent walkers or cyclists pre-pandemic expecting more change than those who were not. Overall, more than 20% identify taking more walks as one of the top three aspects of pandemic life they enjoy.

Many cities have provided temporary infrastructure for walking and biking during the pandemic (13). To support a long-term shift, cities could make these changes permanent. Since commuting traffic is not expected to fully rebound, there is an opportunity to reallocate underutilized road space to pedestrians and bicyclists.

**Urban exodus?**

Some observers project a long-term decline of city centers, as urbanites seek more space and no longer need to commute as often (14). Other research indicates the pandemic has not led longtime urbanites to leave cities (15).

We compare reasons for moving between those who moved from dense urban neighborhoods and all other movers during the first seven months of the pandemic. The main difference was in the extent to which telecommuting opportunities motivated their moves. More than 20% of dense urban employed movers cite not needing to commute as a reason for their move, as opposed to 9% of other employed movers. Likewise, 40% of dense urban employed movers expect to telecommute at least a few times per week post-pandemic, compared to 27% of other employed movers.

Notably, dense urban movers were *not* more likely than other movers to be motivated by either pandemic-related public health concerns or by a desire for a more comfortable home.

The COVID Future dataset strongly suggests that society should expect and be planning for a "new normal." Although only time will reveal the true impact of the pandemic, these data reflect our collective expectations of what the future will bring, providing important insights to help plan for what's next.

**Materials and Methods**

The COVID Future survey dataset that is the basis for this article was collected between July and October 2020. The study protocol was approved by Institutional Review Boards at both Arizona State University and the University of Illinois at Chicago. Online consent was obtained from all survey respondents.

The data are weighted to represent the U.S. population along the dimensions of gender, age, educational attainment, Hispanic status, income, vehicle ownership, and presence of children. All analysis presented here used these weights. A complete description of this dataset is available (2), and both the dataset and the survey questionnaire are available for download (14). The supplementary materials provide details for all calculations.

**Acknowledgments**

Thanks first and foremost to the 7,613 respondents to the COVID Future survey who made this research possible. Thanks also to all who participated in the initial survey design brainstorming meeting in March 2020, as well as key survey pretesters Pat Mokhtarian and Yongsung Lee, who provided suggestions to improve the survey instrument. This research was supported in part by three National Science Foundation grants, the U.S. Department of Transportation, the Knowledge Exchange for Resilience at Arizona State University and a CONVERGE COVID Working Group. Any opinions, findings, conclusions, or recommendations expressed in this material are those of the authors and do not necessarily reflect the views of the funders.

**Supplementary Text for**

**The potential stickiness of pandemic-induced behavior changes in the United States**

Deborah Salon, Matthew Wigginton Conway, Denise Capasso da Silva, Rishabh Singh Chauhan, Sybil Derrible, Kouros Mohammadian, Sara Khoeini, Nathan Parker, Laura Mirtich, Ali Shamshiripour, Ehsan Rahimi, and Ram Pendyala

This supplement provides details about how we analyzed the COVID Future survey data to arrive at the numbers that we cite in the article. It is written in Markdown so that it incorporates the Stata code used for analysis and can be both understood and replicated easily. All explanations are organized by section of the main text, and quotations from the main text are included so that it is clear which results are being documented.

**Introduction**

The results shared in this section of the article are directly calculated from simple tabulations of the survey data, as below.

- **"… more than 70% of respondents indicated there were aspects of pandemic life they would like to continue."**

The following tabulation illustrates that the exact percentage of the weighted sample that reported that they would definitely or maybe like to continue some aspect of pandemic life is 73.9%. In the survey, this question was followed by a list of possible pandemic-era lifestyle aspects, and respondents were instructed to select up to three of them. The top choices were "Working from home, at least some of the time", "Taking more walks", and "Spending more time with family".

```
. tab enjoychange_num [aw=weight_w1b]
```

| Enjoy aspect pandemic | Freq. | Percent | Cum. |
|---|---|---|---|
| No | 1,987.1256 | 26.10 | 26.10 |
| Maybe | 3,393.2894 | 44.57 | 70.67 |
| Yes | 2,232.585 | 29.33 | 100.00 |
| Total | 7,613 | 100.00 | |

**Remote work and its consequences**

- **"The fraction of workers who expect to work remotely at least a few times each week is double that of the pre-pandemic period, increasing from 13% to 26%."**

The following tabulations illustrate. Note that these tabulations include only those workers who were employed pre-pandemic and expect to be employed post-pandemic.

```
. tab wfh_pre_comb3 if wfh_exp_comb3~=. [aw=weight_w1b]
```

| Pre-pandemic remote work frequency | Freq. | Percent | Cum. |
|---|---|---|---|
| Unable | 2,948.97721 | 66.21 | 66.21 |
| Choose not to | 455.476206 | 10.23 | 76.44 |
| A few times/month | 466.154541 | 10.47 | 86.90 |
| More than once/week | 583.392044 | 13.10 | 100.00 |
| Total | 4,454 | 100.00 | |

```
. tab wfh_exp_comb3 if wfh_pre_comb3~=. [aw=weight_w1b]
```

| Expected post-pandemic remote work frequency | Freq. | Percent | Cum. |
|---|---|---|---|
| Unable | 2,474.28919 | 55.55 | 55.55 |
| Choose not to | 216.891242 | 4.87 | 60.42 |
| A few times/month | 621.641647 | 13.96 | 74.38 |
| More than once/week | 1,141.1779 | 25.62 | 100.00 |
| Total | 4,454 | 100.00 | |

- **"Among those new to working remotely at least a few times a week, two-thirds identified remote work and/or commuting less often as key features of pandemic life they would like to continue into the future."**

The tabulations below indicate the fraction of those new to frequent remote work who value each of these aspects of pandemic life, and also the fraction who value either of them. Because we allowed survey respondents to choose up to three aspects of pandemic life that they value, many who were new to frequent remote work chose both of these. We identify those new to frequent remote work as workers who did not work remotely at least a few times a week pre-pandemic, and who do work remotely at least 2 times per week during the pandemic.

```
. tab enjoy_wfh_all [aw=weight_w1b] if wfh_now_days>1 & wfh_now_days~=. & wfh_
pre_comb3<3
```

|             Enjoy remote work |       Freq. |   Percent |    Cum. |
|------------------------------:|------------:|----------:|--------:|
|              Not particularly |  425.009273 |     39.72 |   39.72 |
| Among my top 3 pandemic activities |  644.990727 |     60.28 |  100.00 |
|                         Total |       1,070 |    100.00 |         |

```
. tab enjoy_commute_less_all [aw=weight_w1b] if wfh_now_days>1 & wfh_now_days~
=. & wfh_pre_comb3<3
```

|           Enjoy commuting less |       Freq. |   Percent |    Cum. |
|-------------------------------:|------------:|----------:|--------:|
|               Not particularly |  762.331253 |     71.25 |   71.25 |
| Among my top 3 pandemic activities |  307.668747 |     28.75 |  100.00 |
|                          Total |       1,070 |    100.00 |         |

```
. tab enjoy_both [aw=weight_w1b] if wfh_now_days>1 & wfh_now_days~=. & wfh_pre
_comb3<3
```

| Enjoy remote work OR commuting less |       Freq. |   Percent |    Cum. |
|------------------------------------:|------------:|----------:|--------:|
|                    Not particularly |  350.406814 |     32.75 |   32.75 |
| Among my top 3 pandemic activities |  719.593186 |     67.25 |  100.00 |
|                               Total |       1,070 |    100.00 |         |

- **"More than 70% of those new to regular remote work report that their productivity has stayed the same or improved during the pandemic."**

Note in the tabulation below that the "decreased" productivity categories added together sum to 27.7%. There is another category selected by 8.7% of those new to remote work: "in some ways it has increased and in other ways it has decreased". Because there are both effects for these workers, we included them in the "stayed the same" category in our reporting.

```
. tab prod_change [aw=weight_w1b] if wfh_now_days>1 & wfh_now_days~=. & wfh_pr
e_comb3<3
```

|           Change in work productivity |        Freq. |   Percent |    Cum. |
|--------------------------------------:|-------------:|----------:|--------:|
|                        About the same |   278.620595 |     26.04 |   26.04 |
|              Decreased significantly  |   50.7555587 |      4.74 |   30.78 |
|                  Decreased somewhat   |   245.796212 |     22.97 |   53.75 |
|   In some ways it has increased and in .. |   92.8835689 |      8.68 |   62.44 |
|              Increased significantly  |  107.7849548 |     10.07 |   72.51 |
|                   Increased somewhat  |   294.159111 |     27.49 |  100.00 |
|                                 Total |        1,070 |    100.00 |         |

- **"Among workers who were not frequent telecommuters pre-pandemic, those who hold a bachelor's degree or live in households earning over $100,000 per year are twice as likely to expect to telecommute at least a few times a week post-pandemic."**

```
. tab freq_wfh_change bach [aw=weight_w1b], col nofreq

    Expected  |
post-pandemic |
change in WFH |         bach
    Frequency | No Bachel   Bachelor |     Total
--------------+----------------------+----------
Stopped Freq WFH |    1.52       2.56 |      1.92
  Never Freq WFH |   80.55      59.49 |     72.46
 Always Freq WFH |    8.31      15.77 |     11.18
Started Freq WFH |    9.62      22.17 |     14.44
--------------+----------------------+----------
          Total |  100.00     100.00 |    100.00

. tab freq_wfh_change inc_over100K [aw=weight_w1b], col nofreq

    Expected  |
post-pandemic |
change in WFH |      inc_over100K
    Frequency |   <$100K      >$100K |     Total
--------------+----------------------+----------
Stopped Freq WFH |    1.95       1.89 |      1.92
  Never Freq WFH |   78.84      64.85 |     72.46
 Always Freq WFH |    9.25      13.48 |     11.18
Started Freq WFH |    9.97      19.78 |     14.44
--------------+----------------------+----------
          Total |  100.00     100.00 |    100.00
```

A limitation of the survey in the commuting section is important to mention. Those respondents who were not employed during the pandemic period were not asked what they expect their commute mode to be in the post-pandemic period. In addition, those who commuted for both employment and to school were only asked to provide details on the commute that was their primary activity. In all commuting analyses in this article, therefore, only those survey respondents who were employed in both the pre-pandemic period and the current period, and for whom work was their primary activity, are included.

- **"…we estimate that less frequent commuting will reduce car commute kilometers by approximately 15%."**

The total car commute distance decline is the net result of four effects:

1. Pre-pandemic car commuters expecting to switch away from cars.
2. Pre-pandemic non-car commuters expecting to switch to cars.
3. Pre-pandemic car commuters expecting to increase their remote work frequency without switching modes.
4. Pre-pandemic car commuters expecting to decrease their remote work frequency without switching modes.

To estimate the total change in car commute kilometers, we estimated car commute distances in both the pre- and post-pandemic periods for those survey respondents who were commuting to work in both periods, and calculated the percent change. To be sure that the change was largely due to less frequent commuting rather than mode switching, we also separately calculated the changes due to items (1), (2), and (3+4) above. The total percent change in car commute distance was 14.5%, which decomposed into a 15% decrease in car commute distance from changes in remote work frequency, and a 0.5% increase in car commute distance from mode switching.

To accomplish this calculation with the COVID Future survey data, however, a number of steps and some assumptions were required.

First, we needed to impute commute distances for the 285 pre-pandemic car commuters who did not provide them, but did provide commute times. To do this, we first calculated the average car commute speed for respondents who reported both car commute distances and times, and used this average speed to estimate commute distances for those car commuters who only reported commute times.

Calculating the average weekly car commute kilometers for private car commuters in the pre-pandemic period is straightforward: multiply the reported commute distance by the reported number of days per week of commuting.

For the post-pandemic period, the survey data do not include the number of days per week that each person expects to be commuting. The data do include both the pre-pandemic frequency of remote work and the expected post-pandemic frequency of remote work, however, which can be used together with the pre-pandemic number of days commuted per week to estimate post-pandemic commute days per week. Multiplying this by the reported commute distance gives us the post-pandemic expected weekly car commute kilometers.

Specifically, we estimate the post-pandemic commute frequency as follows.

5. If the pre- and post-pandemic frequency of remote work is the same, then we assume the post-pandemic commute frequency is the pre-pandemic number of days per week.
6. If the pre-pandemic frequency of remote work is "never" or "few times/year" and the post-pandemic frequency of remote work is "once/week" or "few times/month", then we assume the post-pandemic commute frequency is the pre-pandemic number of days per week *minus one*.
7. If the pre-pandemic frequency of remote work is "never" or "few times/year" and the post-pandemic frequency of remote work is "few times/week", then we assume the post-pandemic commute frequency is the pre-pandemic number of days per week *divided by two*.
8. If the pre-pandemic frequency of remote work is "once/week" or "few times/month" and the post-pandemic frequency of remote work is "few times/week", then we assume the post-pandemic commute frequency is the pre-pandemic number of days per week *plus one, divided by two*.
9. If the pre-pandemic frequency of remote work is "once/week" or "few times/month" and the post-pandemic frequency of remote work is "never" or "few times/year" then we assume the post-pandemic commute frequency is the pre-pandemic number of days per week *plus one*.
10. If the pre-pandemic frequency of remote work is "few times/week" and the post-pandemic frequency of remote work is "never" or "few times/year" then we assume the post-pandemic commute frequency is the pre-pandemic number of days per week *multiplied by two*.
11. If the pre-pandemic frequency of remote work is "few times/week" and the post-pandemic frequency of remote work is "once/week" or "few times/month", then we assume the post-pandemic commute frequency is the pre-pandemic number of days per week *minus one, multiplied by two*.
12. If the post-pandemic frequency of remote work is "every day", then we assume the post-pandemic commute frequency is *zero*.
13. If the reported post-pandemic primary mode choice is "I expect to work only from home and not commute", then we assume the post-pandemic commute frequency is *zero*.
14. If the pre-pandemic reported frequency was *zero* because the person worked exclusively from home, but they expect to commute post-pandemic, then we add 5 commute days per week for those who expect not to work remotely, 4 commute days per week for those who expect to work remotely "once/week" or "few times/month", and 2.5 commute days per week for those who expect to work remotely "few times/week".
15. Finally, we adjust so that any resulting post-pandemic commute frequency that has been estimated to be less than zero is reset to zero, and any that has been estimated to be greater than seven is reset to seven. There are a small number of observations in each category.

The resulting average weekly car commute kilometers for the pre- and post-pandemic periods are illustrated below.

First, we calculate the total change in car commute distance for all of those who said that they commuted by car in either pre- or post-pandemic period AND who had non-missing numbers of commute days in both periods AND that had less than 500 car commute miles per week in both periods.

```
. quietly sum pre_car_commute_miles_week [aw=weight_w1b] if
    pre_work_com_days~=. &
    exp_work_com_days~=. &
    pre_work_pri_mode_w1b~="Question not displayed to respondent" &
    wcom_mode_exp_w1b~="Question not displayed to respondent" &
    wcom_mode_exp_w1b~="I do not expect to be employed in the future" &
    pre_car_commute_miles_week<500 &
    exp_car_commute_miles_week<500

. scalar pre_car_miles_total=r(sum)

. quietly sum exp_car_commute_miles_week [aw=weight_w1b] if
    pre_work_com_days~=. &
    exp_work_com_days~=. &
    pre_work_pri_mode_w1b~="Question not displayed to respondent" &
    wcom_mode_exp_w1b~="I do not expect to be employed in the future" &
    wcom_mode_exp_w1b~="Question not displayed to respondent" &
    pre_car_commute_miles_week<500 &
    exp_car_commute_miles_week<500

. scalar exp_car_miles_total=r(sum)

. scalar carmiles_change_total=pre_car_miles_total-exp_car_miles_total

. scalar pct_change_carmiles=(carmiles_change_total)/pre_car_miles_total

. display "Total percent change in car commute distance per week = " pct_chang
e_carmiles
Total percent change in car commute distance per week = .14544338
```

Next, we estimate the contributions to this net decrease in car commute distance that come from mode switching and changes in remote work frequency.

```
. *First, the contribution from those who shift away from cars
. quietly sum pre_car_commute_miles_week [aw=weight_w1b] if
      pre_work_pri_mode_w1b=="Private vehicle" &
      wcom_mode_exp_w1b~="Private vehicle" &
      wcom_mode_exp_w1b~="I expect to work only from home and not commute" &
      pre_work_com_days~=. &
      exp_work_com_days~=. &
      pre_work_pri_mode_w1b~="Question not displayed to respondent" &
      wcom_mode_exp_w1b~="Question not displayed to respondent" &
      wcom_mode_exp_w1b~="I do not expect to be employed in the future" &
      pre_car_commute_miles_week<500 &
      exp_car_commute_miles_week<500

. scalar carmiles_change_shift_away=r(sum)

. *Second, the contribution from those who shift to cars
. quietly sum exp_car_commute_miles_week [aw=weight_w1b] if
      pre_work_pri_mode_w1b~="Private vehicle" &
      wcom_mode_exp_w1b=="Private vehicle" &
      pre_work_com_days~=. &
      exp_work_com_days~=. &
      pre_work_pri_mode_w1b~="Question not displayed to respondent" &
      wcom_mode_exp_w1b~="Question not displayed to respondent" &
      wcom_mode_exp_w1b~="I do not expect to be employed in the future" &
      pre_car_commute_miles_week<500 &
      exp_car_commute_miles_week<500

. scalar carmiles_change_shift_to=r(sum)

. *Finally, the contribution from those pre-pandemic car commuters
. *who change their frequency of remote work
. quietly sum pre_car_commute_miles_week [aw=weight_w1b] if
      pre_work_pri_mode_w1b=="Private vehicle" &
      (wcom_mode_exp_w1b=="Private vehicle" |
      wcom_mode_exp_w1b=="I expect to work only from home and not commute") &
      pre_work_com_days~=. &
      exp_work_com_days~=. &
      pre_work_pri_mode_w1b~="Question not displayed to respondent" &
      wcom_mode_exp_w1b~="Question not displayed to respondent" &
      wcom_mode_exp_w1b~="I do not expect to be employed in the future" &
      pre_car_commute_miles_week<500 &
      exp_car_commute_miles_week<500

. scalar pre_carmiles_week3=r(sum)

. quietly sum exp_car_commute_miles_week [aw=weight_w1b] if
      pre_work_pri_mode_w1b=="Private vehicle" &
      (wcom_mode_exp_w1b=="Private vehicle" |
      wcom_mode_exp_w1b=="I expect to work only from home and not commute") &
      pre_work_com_days~=. &
```

```
        exp_work_com_days~=. &
        pre_work_pri_mode_w1b~="Question not displayed to respondent" &
        wcom_mode_exp_w1b~="Question not displayed to respondent" &
        wcom_mode_exp_w1b~="I do not expect to be employed in the future" &
        pre_car_commute_miles_week<500 &
        exp_car_commute_miles_week<500

. scalar exp_carmiles_week3=r(sum)

. scalar carmiles_change_wfh=pre_carmiles_week3-exp_carmiles_week3

. scalar pct_carmiles_change_wfh=carmiles_change_wfh/carmiles_change_total

. display "Fraction of change in car commute distance per week due to changes
in remote work = " pct_carmiles_change_wfh
Fraction of change in car commute distance per week due to changes in remote w
ork = 1.0302229
```

Importantly, in both this and the analysis of transit commute impacts of COVID, we assume that if people are working remotely on a given day, then they are not commuting on that day. Because the COVID Future survey is focused on impacts of the COVID-19 pandemic, we believe that this assumption is valid. A major impact of the pandemic has been that many workers have switched from working at a workplace to working from their homes. Therefore, the pandemic context clearly suggests that respondents to the COVID Future survey likely would interpret questions about remote work frequency to mean remote work *instead of* commuting to a workplace, rather than remote work *in addition to* commuting to a workplace.

In both analyses, we further assume that commuters use only their reported primary commute mode on the days when they travel to their workplace.

- **"The fraction of commuters who choose the car as their primary commute mode is not expected to change substantially."**

The following tabulations illustrate that 86% of U.S. workers were car commuters pre-pandemic, and this figure is expected to be 85% post-pandemic.

```
. tab pre_work_pri_mode_w1b [aw=weight_w1b] if
        pre_work_pri_mode_w1b~="Question not displayed to respondent" &
        wcom_mode_exp_w1b~="Question not displayed to respondent" &
        wcom_mode_exp_w1b~="I do not expect to be employed in the future"
```

| Pre-pandemic primary commute mode | Freq. | Percent | Cum. |
|---:|---:|---:|---:|
| Other mode | 39.3251632 | 1.22 | 1.22 |
| Personal bicycle/scooter | 33.7477706 | 1.05 | 2.27 |
| Private vehicle | 2,776.3097 | 86.30 | 88.57 |
| Shared bicycle/scooter | 14.962213 | 0.47 | 89.04 |
| Transit | 267.030076 | 8.30 | 97.34 |
| Walk | 85.6250465 | 2.66 | 100.00 |
| Total | 3,217 | 100.00 | |

```
. tab wcom_mode_exp_w1b [aw=weight_w1b] if
    wcom_mode_exp_w1b~="Question not displayed to respondent" &
    wcom_mode_exp_w1b~="I do not expect to be employed in the future" &
    pre_work_pri_mode_w1b~="Question not displayed to respondent"
```

| Expected post-pandemic primary commute mode | Freq. | Percent | Cum. |
|---|---|---|---|
| I expect to work only from home and n.. | 117.837023 | 3.66 | 3.66 |
| Other mode | 22.2440893 | 0.69 | 4.35 |
| Personal bicycle/scooter | 49.936278 | 1.55 | 5.91 |
| Private vehicle | 2,730.1952 | 84.87 | 90.77 |
| Shared bicycle/scooter | 14.9091232 | 0.46 | 91.24 |
| Transit | 204.861871 | 6.37 | 97.61 |
| Walk | 77.0163873 | 2.39 | 100.00 |
| Total | 3,217 | 100.00 | |

- **"Commuting accounted for about half of all transit trips pre-pandemic."**

For comparison purposes, we determined the percentage of transit trips that were for commute purposes pre-COVID using the 2017 National Household Travel Survey. We subsetted the trips file to only transit trips (TRPTRANS codes 11 public/commuter bus, 12 paratransit, 15 Amtrak/commuter rail, and 16 Subway/elevated/light rail). We did not include boat trips as they may include both ferries and privately-operated boats, and excluded trips where the trip purpose was not included. We computed the weighted proportion of these trips which were for commuting to work or school purposes (either to work or from work).

- **"Our data suggest nearly a 40% decline in transit commute trips post-pandemic, relative to pre-pandemic."**

In order to arrive at this result, we use the assumptions outlined above to estimate the post-pandemic commute frequency per week. We then calculate the total number of commute trips per week by transit across all commuters for the pre- and post-pandemic periods, and compute the percent change.

```
. quietly sum pre_work_com_days [aw=weight_w1b] if
    pre_work_pri_mode_w1b=="Transit" &
    pre_work_com_days~=. &
    exp_work_com_days~=. &
    pre_work_pri_mode_w1b~="Question not displayed to respondent" &
    wcom_mode_exp_w1b~="I do not expect to be employed in the future" &
    wcom_mode_exp_w1b~="Question not displayed to respondent"

. scalar pre_trans_trips_week=r(sum)

. quietly sum exp_work_com_days [aw=weight_w1b] if
    wcom_mode_exp_w1b=="Transit" &
    pre_work_com_days~=. &
    exp_work_com_days~=. &
    pre_work_pri_mode_w1b~="Question not displayed to respondent" &
    wcom_mode_exp_w1b~="I do not expect to be employed in the future" &
    wcom_mode_exp_w1b~="Question not displayed to respondent"
```

```
. scalar exp_trans_trips_week=r(sum)

. scalar trans_change_total=pre_trans_trips_week-exp_trans_trips_week

. scalar pct_change_transit=(trans_change_total)/pre_trans_trips_week

. display "Total percent change in transit commute days per week = " pct_chang
e_transit
Total percent change in transit commute days per week = .38953326
```

- **"About half of this decline can be attributed to changes in the frequency of remote work, while the remainder comes from commute mode shifts."**

The analysis behind this statement is exactly analogous to that which we conducted for car commute distance. For transit, we do not calculate the change in distance commuted by transit but instead focus only on the change in the number of transit trips.

The total transit commute trip decline is the net result of four effects:

1. Pre-pandemic transit commuters expecting to switch away from transit.
2. Pre-pandemic non-transit commuters expecting to switch to transit.
3. Pre-pandemic transit commuters expecting to increase their remote work frequency without switching modes.
4. Pre-pandemic transit commuters expecting to decrease their remote work frequency without switching modes.

Here, we estimate the portion of the total transit commute trip decline that is due to the last two of these. For completeness, the following Stata code calculates the transit commute trip changes from expected mode shifts as well as that from changes in the frequency of remote work. Together these changes equal the total change in transit demand.

```
. *First, the contribution from those who shift away from transit
. quietly sum pre_work_com_days [aw=weight_w1b] if
    pre_work_pri_mode_w1b=="Transit" &
    wcom_mode_exp_w1b~="Transit" &
    wcom_mode_exp_w1b~="I expect to work only from home and not commute" &
    pre_work_com_days~=. &
    exp_work_com_days~=. &
    pre_work_pri_mode_w1b~="Question not displayed to respondent" &
    wcom_mode_exp_w1b~="I do not expect to be employed in the future" &
    wcom_mode_exp_w1b~="Question not displayed to respondent"

. scalar trans_trips_change_shift_away=r(sum)

. *Second, the contribution from those who shift to transit
. quietly sum exp_work_com_days [aw=weight_w1b] if
    pre_work_pri_mode_w1b~="Transit" &
    wcom_mode_exp_w1b=="Transit" &
    pre_work_com_days~=. &
    exp_work_com_days~=. &
    pre_work_pri_mode_w1b~="Question not displayed to respondent" &
    wcom_mode_exp_w1b~="I do not expect to be employed in the future" &
    wcom_mode_exp_w1b~="Question not displayed to respondent"
```

```
. scalar trans_trips_change_shift_to=r(sum)

. *Third, the contribution from changes in remote work frequency
. *This group are transit users in both periods, so calculate both periods
. *and take the difference
. quietly sum pre_work_com_days [aw=weight_w1b] if
    pre_work_pri_mode_w1b=="Transit" &
    (wcom_mode_exp_w1b=="Transit" |
    wcom_mode_exp_w1b=="I expect to work only from home and not commute") &
    pre_work_com_days~=. &
    exp_work_com_days~=. &
    pre_work_pri_mode_w1b~="Question not displayed to respondent" &
    wcom_mode_exp_w1b~="I do not expect to be employed in the future" &
    wcom_mode_exp_w1b~="Question not displayed to respondent"

. scalar pre_trans_trips_week3=r(sum)

. quietly sum exp_work_com_days [aw=weight_w1b] if
    pre_work_pri_mode_w1b=="Transit" &
    (wcom_mode_exp_w1b=="Transit" |
    wcom_mode_exp_w1b=="I expect to work only from home and not commute") &
    pre_work_com_days~=. &
    exp_work_com_days~=. &
    pre_work_pri_mode_w1b~="Question not displayed to respondent" &
    wcom_mode_exp_w1b~="I do not expect to be employed in the future" &
    wcom_mode_exp_w1b~="Question not displayed to respondent"

. scalar exp_trans_trips_week3=r(sum)

. scalar trans_trips_change_wfh=pre_trans_trips_week3-exp_trans_trips_week3

. scalar pct_trans_change_wfh=trans_trips_change_wfh/trans_change_total

. display "Fraction of change in transit commute days per week due to changes in
remote work = " pct_trans_change_wfh
Fraction of change in transit commute days per week due to changes in remote work = .48711312
```

- **"Of this decline, about half can be attributed to changes in commuting frequency, 40% comes from a net shift among transit commuters toward the private car, and the remaining 10% comes from shifts to other modes."**

The first part of this sentence is calculated above at 48.7%. To calculate the fraction that comes from a net shift toward the private car versus other modes, the following additional code is required.

```
. quietly sum pre_work_com_days [aw=weight_w1b] if
    pre_work_pri_mode_w1b=="Transit" &
    wcom_mode_exp_w1b=="Private vehicle" &
    wcom_mode_exp_w1b~="I expect to work only from home and not commute" &
    pre_work_com_days~=. &
    exp_work_com_days~=. &
    pre_work_pri_mode_w1b~="Question not displayed to respondent" &
```

```
        wcom_mode_exp_w1b~="I do not expect to be employed in the future" &
        wcom_mode_exp_w1b~="Question not displayed to respondent"

. scalar pretrans_to_car=r(sum)

. quietly sum exp_work_com_days [aw=weight_w1b] if
        pre_work_pri_mode_w1b=="Private vehicle" &
        wcom_mode_exp_w1b=="Transit" &
        pre_work_com_days~=. &
        exp_work_com_days~=. &
        pre_work_pri_mode_w1b~="Question not displayed to respondent" &
        wcom_mode_exp_w1b~="I do not expect to be employed in the future" &
        wcom_mode_exp_w1b~="Question not displayed to respondent"

. scalar precar_to_trans=r(sum)

. scalar pct_tofrom_car=(pretrans_to_car - precar_to_trans)/trans_change_total

. disp "Percent change in transit commutes/week from shift to/from private car
 = " round(pct_tofrom_car*100) "%"
Percent change in transit commutes/week from shift to/from private car = 41%
```

- **"Our data suggest that the number of people who plan to dine in restaurants at least a few times each week will decrease by more than 20% post-pandemic, compared to the pre-pandemic era."**

This result is derived from the data in the following tabulations.

```
. tab shdi_pre_4 [aw=weight_w1b]
```

| Pre-pandemic restaurant dining frequency | Freq. | Percent | Cum. |
|---|---|---|---|
| Never | 352.603441 | 4.63 | 4.63 |
| A few times/year | 1,185.9179 | 15.58 | 20.21 |
| A few times/month | 3,628.6267 | 47.66 | 67.87 |
| A few times/week | 2,321.8066 | 30.50 | 98.37 |
| Every day | 124.04527 | 1.63 | 100.00 |
| Total | 7,613 | 100.00 | |

```
. tab shdi_exp_restaurant_dinein_w1b [aw=weight_w1b]
```

|       Expected post-pandemic restaurant dining frequency | Freq. | Percent | Cum. |
|---|---|---|---|
| Never           | 422.711151   | 5.55   | 5.55   |
| A few times/year | 1,505.2189  | 19.77  | 25.32  |
| A few times/month | 3,761.4631 | 49.41  | 74.73  |
| A few times/week | 1,791.3902  | 23.53  | 98.26  |
| Every day       | 132.216646   | 1.74   | 100.00 |
| Total           | 7,613        | 100.00 |        |

Before the pandemic = 2,321.81+ 124.04= 2,445.85

After the pandemic = 1,791.39 + 132.22= 1,923.61

Ratio = 1,923.61/2,445.85 = 0.7867

- **"…the restaurant industry alone employed 8% of American workers pre-pandemic…"**

This figure is calculated by dividing the employment in the "food services and drinking places" industry (12.308 million) by total non-farm employment in the U.S. (152.523 million) for February 2020.

**Figure 1**

Figure 1 illustrates how the pandemic is expected to change demand for remote work and car commuting. The data behind this chart come from the following tabulations. Car commuting frequency here is calculated more simply than described above. Specifically, those who "Always" car commute report a private vehicle as their primary commute mode and that they work remotely "Never" or only "Few times/year". Those who car commute "Most days" are car commuters who work remotely "Few times/month" or "Once/week". Those who car commute "Some days" are car commuters who work remotely "Few times/week". Finally, those who car commute "Never" either work remotely "Every day" or commute using another transport mode.

As explained above, the commuting analysis is limited to those who answered the survey questions about commute mode for both the pre-pandemic and post-pandemic periods.

```
. tab wfh_pre_comb3 [aw=weight_w1b]
```

| Pre-pandemic remote work frequency | Freq. | Percent | Cum. |
|---|---|---|---|
| Unable             | 2,948.97721 | 66.21  | 66.21  |
| Choose not to      | 455.476206  | 10.23  | 76.44  |
| A few times/month  | 466.154541  | 10.47  | 86.90  |
| More than once/week | 583.392044 | 13.10  | 100.00 |
| Total              | 4,454       | 100.00 |        |

```
. tab wfh_exp_comb3 [aw=weight_w1b]
```

| Expected post-pandemic remote work frequency | Freq. | Percent | Cum. |
|---|---|---|---|
| Unable | 2,497.3628 | 55.03 | 55.03 |
| Choose not to | 226.099571 | 4.98 | 60.01 |
| A few times/month | 628.921281 | 13.86 | 73.87 |
| More than once/week | 1,185.6163 | 26.13 | 100.00 |
| Total | 4,538 | 100.00 | |

```
. tab pre_car_comm_freq [aw=weight_w1b] if exp_car_comm_freq~=.
```

| Pre-pandemic car commute frequency | Freq. | Percent | Cum. |
|---|---|---|---|
| Always | 2,210.0229 | 68.70 | 68.70 |
| Most days | 313.525703 | 9.75 | 78.44 |
| Some days | 252.761104 | 7.86 | 86.30 |
| Never | 440.690269 | 13.70 | 100.00 |
| Total | 3,217 | 100.00 | |

```
. tab exp_car_comm_freq [aw=weight_w1b] if pre_car_comm_freq~=.
```

| Expected post-pandemic car commute frequency | Freq. | Percent | Cum. |
|---|---|---|---|
| Always | 1,743.6142 | 54.20 | 54.20 |
| Most days | 417.409609 | 12.98 | 67.18 |
| Some days | 569.171418 | 17.69 | 84.87 |
| Never | 486.804772 | 15.13 | 100.00 |
| Total | 3,217 | 100.00 | |

**Figure 2**

Figure 2 illustrates how the pandemic is expected to change demand for both air travel and non-motorized mode use. The survey questions that form the basis for the figure included questions about the frequency of engaging in these activities pre-pandemic, how that frequency will change post-pandemic, and, for changes in expected air travel and bicycling frequency, we asked why. Specifically, the survey questions were:

- "How much do you expect your airplane travel for leisure/personal (business) purposes to change once COVID-19 is no longer a threat, compared to your level of travel before the COVID-19 pandemic?"
- "Why do you anticipate an increase/decrease in your long-distance travel for leisure/personal (business) purposes after COVID-19 is no longer a threat? Select all that apply." (separate questions for increase and decrease, depending on the person's actual response to the previous question)
- "After COVID-19 is no longer a threat, how do you expect your use of the following means of transport to change, relative to before the COVID-19 pandemic?" (The prompt also included the text "Please include any walks or bike rides for exercise or enjoyment.")
- "Why do you expect to increase your use of bicycles? Please select all that apply."

The air travel portion of the Figure focuses only on those who had traveled by airplane at least once per year pre-pandemic for leisure/personal and, separately, for business purposes (based on a separate question about pre-pandemic air travel). Those who expect to increase/decrease air travel for each purpose were then asked to select the reasons why, which we separated into "Pandemic-related", "New realization", and "Other" categories, as follows:

- Leisure Only, Less, new realization: "I am able to use technology (e.g., FaceTime, Zoom) to meaningfully engage with long-distance connections"
- Business Only, Less, new realization: "I realized I could conduct my meetings by conference call/video conference"
- Business Only, Less, new realization: "Those I meet with have realized that we can conduct meetings by conference call/video conference"
- Leisure and Business, Less, new realization: "I want to spend more time at home"
- Leisure and Business, Less, other reason: "I anticipate taking more of my long-distance trips by car"
- Leisure and Business, Less, other reason: "I anticipate taking more of my long-distance trips by train or bus"
- Leisure and Business, Less, other reason: "I want to fly less for environmental reasons"
- Business Only, Less, other reason: "My employer adopted a commitment to reduce travel by airplane"
- Business Only, Less, other reason: "My job responsibilities have changed"
- Business Only, Less, other reason: "I expect reduced budget for travel"
- Leisure and Business, Less, other reason: "Other, please specify"
- Leisure and Business, Less, pandemic-related: "I will not feel safe or comfortable sharing close space with strangers"
- Leisure Only, Less, pandemic-related: "My financial circumstances changed and I can no longer afford to travel in the same way"
- Leisure and Business, More, pandemic-related: "I will need/want to take trips that were cancelled during the COVID-19 pandemic"
- Leisure Only, More, pandemic-related: "After having been cooped up at home for so long, I want to travel more than I did before"
- Leisure Only, More, other reason: "My financial circumstances changed and I can now afford more air travel"
- Business Only, More, other reason: "My job responsibilities have changed"
- Leisure and Business, More, other reason: "Other, please specify"

If a respondent selected both a "New realization" reason and an "Other" reason, their response was categorized as a "New realization" response. If a respondent selected both an "Other" reason and a "Pandemic-related" reason, their response was categorized as an "Other" response. Therefore, those categorized as "Pandemic-related" are those who *only* selected a "Pandemic-related" reason.

With these definitions, below are the tabulations of the proportion of the sample in each category for personal and business air travelers that are illustrated in Figure 2.

```
. tab ld_bz_exp_reason [aw=weight_w1b]
```

| Expected post-pandemic business air travel w/reasons | Freq. | Percent | Cum. |
|---|---|---|---|
| Less, new realization | 452.464562 | 27.01 | 27.01 |
| Less, other reason | 197.081616 | 11.77 | 38.78 |
| Less, pandemic-related | 38.2287998 | 2.28 | 41.06 |
| About the same | 749.003526 | 44.72 | 85.78 |
| More, pandemic-related | 129.11137 | 7.71 | 93.49 |
| More, other reason | 109.110126 | 6.51 | 100.00 |
| Total | 1,675 | 100.00 | |

```
. tab ld_per_exp_reason [aw=weight_w1b]
```

| Expected post-pandemic personal air travel w/reasons | Freq. | Percent | Cum. |
|---|---|---|---|
| Less, new realization | 413.306654 | 7.78 | 7.78 |
| Less, other reason | 641.044088 | 12.07 | 19.84 |
| Less, pandemic-related | 856.78342 | 16.13 | 35.97 |
| About the same | 2,666.7692 | 50.19 | 86.16 |
| More, pandemic-related | 497.236891 | 9.36 | 95.52 |
| More, other reason | 237.859707 | 4.48 | 100.00 |
| Total | 5,313 | 100.00 | |

The walking portion of Figure 2 is straightforward, since there was not a survey question that asked about reasons for expected increases or decreases in walking frequency post-pandemic. Below is the tabulation of the proportion of the sample in each category for walking. Figure 2 puts the "Somewhat" and "Much" categories together for both "More" and "Less" walking.

```
. tab tr_freq_exp_walk_w1b [aw=weight_w1b] if tr_freq_exp_walk_w1b~="Seen but unanswered"
```

| Expected post-pandemic walking frequency | Freq. | Percent | Cum. |
|---|---|---|---|
| About the same | 4,803.8198 | 63.53 | 63.53 |
| Much less than before | 253.149255 | 3.35 | 66.87 |
| Much more than before | 707.520755 | 9.36 | 76.23 |
| Somewhat less than before | 280.23573 | 3.71 | 79.94 |
| Somewhat more than before | 1,517.2744 | 20.06 | 100.00 |
| Total | 7,562 | 100.00 | |

Figure 2 also separates those who were Former Regular Users (FRU) of walking from those who were not, with a regular user defined as someone who walked a few times a week or more pre-pandemic.

```
. tab tr_freq_exp_walk_w1b [aw=weight_w1b] if tr_freq_exp_walk_w1b~="Seen but
unanswered" & (tr_freq_pre_walk_w1b==3 | tr_freq_pre_walk_w1b==4)
```

| Expected post-pandemic walking frequency | Freq. | Percent | Cum. |
|---|---|---|---|
| About the same | 2,037.98047 | 54.35 | 54.35 |
| Much less than before | 121.013992 | 3.23 | 57.57 |
| Much more than before | 527.477842 | 14.07 | 71.64 |
| Somewhat less than before | 142.198825 | 3.79 | 75.43 |
| Somewhat more than before | 921.32887 | 24.57 | 100.00 |
| Total | 3,750 | 100.00 | |

```
. tab tr_freq_exp_walk_w1b [aw=weight_w1b] if tr_freq_exp_walk_w1b~="Seen but
unanswered" & tr_freq_pre_walk_w1b<3
```

| Expected post-pandemic walking frequency | Freq. | Percent | Cum. |
|---|---|---|---|
| About the same | 2,740.7571 | 72.24 | 72.24 |
| Much less than before | 125.372119 | 3.30 | 75.54 |
| Much more than before | 188.0367924 | 4.96 | 80.50 |
| Somewhat less than before | 137.971133 | 3.64 | 84.14 |
| Somewhat more than before | 601.862902 | 15.86 | 100.00 |
| Total | 3,794 | 100.00 | |

Calculating the biking portion of Figure 2 was complicated by the fact that the survey asked respondents separately about biking using personal bicycles and biking using shared bicycles, and this figure combines the two. To accomplish this, we assume the following:

- a person expects to bike less post-pandemic than they did pre-pandemic if they expect to use either personal or shared bicycles "much less", and they do not expect to use the other bicycle type "much more"
- a person expects to bike less post-pandemic than they did pre-pandemic if they expect to use either personal or shared bicycles "somewhat less", and they do not expect to use the other bicycle type "somewhat more" or "much more"
- a person expects no change in their biking if they expect to bike "about the same" in both bicycle types
- a person expects no change in their biking if they expect to bike "somewhat less" for one bicycle type and "somewhat more" for the other, or "much less" for one and "much more" for the other
- a person expects to bike more post-pandemic than they did pre-pandemic if they expect to use either personal or shared bicycles "much more", and they do not expect to use the other bicycle type "much less"
- a person expects to bike more post-pandemic than they did pre-pandemic if they expect to use either personal or shared bicycles "somewhat more", and they do not expect to use the other bicycle type "somewhat less" or "much less"

Figure 2 also includes reasons for expected increases in biking, but not reasons for expected decreases. The reasons for expected increases are categorized into "New realization" and "Other" reasons, as follows:

- More, new realization: "I realized I really like biking"
- More, new realization: "I realized biking is fast"
- More, new realization: "I bought a bike"
- More, new realization: "I realized biking is an inexpensive way to get around"
- More, other reason: "I expect my city to make biking safer"
- More, other reason: "I expect to bike more in my neighborhood"
- More, other reason: "I expect to use biking to replace trips by other means of transport"
- More, other reason: "Other, please specify"

If a respondent selected both a "New realization" reason and an "Other" reason, their response was categorized as a "New realization" response.

Finally, Figure 2 illustrates differences in walking and biking expectations for Former Regular Users (FRU) and Former Non-Regular Users (FNRU). As with walking, these are defined as people who used each mode pre-pandemic a few times a week or more. Again, this is somewhat complicated for biking because the survey asked respondents about frequency of bike use separately for personal and shared bicycles. Here, we assume that a person is a FRU if they used either type of bicycle a few times a week or more pre-pandemic, or if they used both types of bicycle a few times a month pre-pandemic.

With these definitions, below are the tabulations of the proportion of the sample overall and in each pre-pandemic user frequency category for biking that are illustrated in Figure 2.

```
. tab bike_exp_reason [aw=weight_w1b]

     Expected biking |
  frequency w/reasons |      Freq.     Percent        Cum.
---------------------+---------------------------------------
                Less |   438.510214        5.93        5.93
      About the same | 5,856.7007         79.20       85.13
   More, other reason|  346.3236583        4.68       89.81
 More, new realization|  753.465462        10.19      100.00
---------------------+---------------------------------------
               Total |      7,395        100.00

. tab bike_exp_reason [aw=weight_w1b] if bike_pre_high==0

     Expected biking |
  frequency w/reasons |      Freq.     Percent        Cum.
---------------------+---------------------------------------
                Less |  286.6207111        4.48        4.48
      About the same | 5,441.5458         85.09       89.57
   More, other reason|  274.027911         4.29       93.86
 More, new realization|  392.805625        6.14      100.00
---------------------+---------------------------------------
               Total |      6,395        100.00
```

```
. tab bike_exp_reason [aw=weight_w1b] if bike_pre_high==1
```

|        Expected biking frequency w/reasons |        Freq. |   Percent |     Cum. |
|---|---|---|---|
|                 Less |   144.090574 |     14.41 |    14.41 |
|       About the same |   446.8948388 |    44.69 |    59.10 |
|    More, other reason |    70.1509578 |     7.02 |    66.11 |
| More, new realization |    338.863629 |    33.89 |   100.00 |
|                Total |        1,000 |   100.00 |          |

**A paradigm shift in air travel**

All of the specific numbers in this section of the article can be derived from the air travel-related information that is represented in Figure 2. Please see Figure 2's explanation for details.

**Accelerated growth of online shopping for groceries**

We identified those who were new to online grocery shopping as people who reported that they "Never" shopped online for grocery delivery or pickup at the store pre-pandemic, and that they did one or both of these activities within the seven-day period before taking the survey during the pandemic. There are undoubtedly others in this sample that also tried online grocery shopping during the pandemic, but did not happen to do so during the week before taking this survey. That said, the subsample that we have identified here as new to online grocery shopping is 780 people - large enough to draw conclusions from.

The remaining results reported in this section of the article are directly calculated from simple tabulations of the survey data, as below.

- **"We analyzed survey responses from those who tried online grocery shopping for the first time during the pandemic. Approximately half expect to continue to grocery shop online at least a few times a month post-pandemic..."**

```
. tab exp_groc_online [aw=weight_w1b] if new_groc_online==1
```

| Expected post-pandemic online grocery shop frequency | Freq. | Percent | Cum. |
|---|---|---|---|
| 0 | 398.346704 | 51.07 | 51.07 |
| 1 | 381.653296 | 48.93 | 100.00 |
| Total | 780 | 100.00 | |

- **"… nearly 90% of them also expect to shop in-store for groceries at least a few times a month."**

```
. tab shdi_exp_groceries_instore_w1b [aw=weight_w1b] if new_groc_online==1
```

|   Expected post-pandemic in-store grocery shop | Freq.      | Percent | Cum.   |
| ---------------------------------------------- | ---------- | ------- | ------ |
| Never                                          | 23.8746581 | 3.06    | 3.06   |
| A few times/year                               | 76.1217172 | 9.76    | 12.82  |
| A few times/month                              | 382.186539 | 49.00   | 61.82  |
| A few times/week                               | 282.392375 | 36.20   | 98.02  |
| Every day                                      | 15.4247104 | 1.98    | 100.00 |
| Total                                          | 780        | 100.00  |        |

- **"This suggests that online grocery shopping does not completely replace in-store shopping, although it may reduce its frequency."**

We investigated whether those new to online grocery shopping also expect to reduce their frequency of in-store shopping, and found that many of them do. Specifically, among respondents who were new to online shopping during the pandemic and also reported that they would still be shopping in person at least a few times each month post-pandemic, 25% of them report that they expect to shop in grocery stores less frequently. This compares to about 10% of respondents who did not try online grocery shopping for the first time during the pandemic. See below for the code that provides these numbers.

```
. tab instore_groc_decr [aw=weight_w1b] if new_groc_online==1 &
instore_pre_groc_cond~=1 & instore_exp_groc_cond~=1
```

| Expected decrease in in-store grocery shop frequency | Freq.      | Percent | Cum.   |
| ---------------------------------------------------- | ---------- | ------- | ------ |
| 0                                                    | 506.632054 | 74.72   | 74.72  |
| 1                                                    | 171.367946 | 25.28   | 100.00 |
| Total                                                | 678        | 100.00  |        |

```
. tab instore_groc_decr [aw=weight_w1b] if new_groc_online==0 &
instore_pre_groc_cond~=1 & instore_exp_groc_cond~=1
```

| Expected decrease in in-store grocery shop frequency | Freq.      | Percent | Cum.   |
| ---------------------------------------------------- | ---------- | ------- | ------ |
| 0                                                    | 5,605.9738 | 88.55   | 88.55  |
| 1                                                    | 725.026185 | 11.45   | 100.00 |
| Total                                                | 6,331      | 100.00  |        |

- **"Among all U.S. residents, 30% expect to grocery shop online at least a few times a month post-pandemic, up from 21% pre-pandemic."**

```
. tab exp_groc_online [aw=weight_w1b]
```

| Expected post-pandemic online grocery shop frequency | Freq. | Percent | Cum. |
|---|---|---|---|
| 0 | 5,366.0266 | 70.49 | 70.49 |
| 1 | 2,246.9734 | 29.51 | 100.00 |
| Total | 7,613 | 100.00 | |

```
. tab pre_groc_online [aw=weight_w1b]
```

| Pre-pandemic online grocery shop frequency | Freq. | Percent | Cum. |
|---|---|---|---|
| 0 | 6,023.0379 | 79.12 | 79.12 |
| 1 | 1,589.9621 | 20.88 | 100.00 |

- **"63% of people expect to shop for durable goods online at least a few times a month post-pandemic, compared to 59% before the pandemic."**

```
. tab shdi_pre_7 [aw=weight_w1b]
```

| Pre-pandemic online non-grocery shop | Freq. | Percent | Cum. |
|---|---|---|---|
| Never | 1,304.4542 | 17.13 | 17.13 |
| A few times/year | 1,848.6911 | 24.28 | 41.42 |
| A few times/month | 3,117.4766 | 40.95 | 82.37 |
| A few times/week | 1,211.4052 | 15.91 | 98.28 |
| Every day | 130.972874 | 1.72 | 100.00 |
| Total | 7,613 | 100.00 | |

```
. tab shdi_exp_onlineother_w1b [aw=weight_w1b]
```

|        Expected post-pandemic online non-grocery shop | Freq. | Percent | Cum. |
|---|---|---|---|
| Never | 1,151.4118 | 15.12 | 15.12 |
| A few times/year | 1,632.4737 | 21.44 | 36.57 |
| A few times/month | 3,235.0604 | 42.49 | 79.06 |
| A few times/week | 1,432.4196 | 18.82 | 97.88 |
| Every day | 161.63451 | 2.12 | 100.00 |
| Total | 7,613 | 100.00 | |

**Marked increases in walking and bicycling**

The results reported in this section of the article are directly calculated from simple tabulations of the survey data, as below.

- **"Post-pandemic, 30% of U.S. residents plan to take walks more frequently post-COVID than they did before the pandemic, and nearly 15% plan to bike more"**
- **"... those who were frequent walkers or cyclists pre-pandemic expecting more change than those who were not."**

The calculations for these statements are documented in the Figure 2 data description above.

- **"More than 20% identify taking more walks as an aspect of pandemic life they enjoy."**

Those survey respondents who responded that there were at least some aspects of pandemic life that they enjoyed were asked to specify up to three, choosing from a list that was generated based on an earlier survey. This tabulation indicates the fraction of the full sample (i.e. including those who do not enjoy any aspects of pandemic life, so representative of the U.S. adult population) that selected "Taking more walks".

```
. tab enjoy_walks_all [aw=weight_w1b]
```

| Enjoy taking more walks | Freq. | Percent | Cum. |
|---|---|---|---|
| Not particularly | 5,926.3723 | 77.85 | 77.85 |
| Among my top 3 pandemic activities | 1,686.6277 | 22.15 | 100.00 |
| Total | 7,613 | 100.00 | |

**Urban exodus?**

For the analysis of urban vs. non-urban movers, zip codes are classified as urban if they have a housing unit density of at least 2000 units per square mile. The results reported below of the article are directly calculated from simple tabulations of the survey data, comparing those movers who previously lived in urban neighborhoods to those who previously lived in lower-density neighborhoods.

- **"More than 20% of dense urban employed movers cite not needing to commute as a reason for their move, as opposed to 9% of other employed movers."**

The tabulation below presents the percent of employed movers coming from dense urban areas who reported that a reason for their move was "I do not need to commute", compared to employed movers coming from lower-density neighborhoods.

```
. tab home_move_why_5_w1b urban_pre [aw=weight_w1b] if home_move_why_5_w1b!="S
een but unanswered" & (worker_pre=="yes" | worker_now=="yes"), col nofreq
```

|  Moving reason: commute | Moved from neighborhood type | | Total |
|---|---|---|---|
|  | Lower-den | Dense urb | |
| I **do not** need to co.. | 8.80 | 21.77 | 11.23 |
| Not selected | 91.20 | 78.23 | 88.77 |
| Total | 100.00 | 100.00 | 100.00 |

- **"Likewise, 40% of dense urban movers expect to work remotely at least a few times per week post-pandemic, compared to 27% of all other movers."**

```
. tab wfh_exp_comb3 urban_pre [aw=weight_w1b], col nofreq
```

| Expected post-pandemic remote work frequency | Moved from neighborhood type | | Total |
|---|---|---|---|
|  | Lower-den | Dense urb | |
| Unable | 48.58 | 32.49 | 45.57 |
| Choose **not** to | 7.05 | 8.57 | 7.33 |
| A few times/month | 17.24 | 18.90 | 17.55 |
| More than once/week | 27.14 | 40.04 | 29.55 |
| Total | 100.00 | 100.00 | 100.00 |

- **"… dense urban movers were not more likely than other movers to be motivated by either pandemic-related public health concerns or by a desire for a more comfortable home."**

Parallel to the tabulation above, the following tabulations present the percent of movers coming from urban areas and from lower-density neighborhoods who reported that a reason for their move was "I did not feel safe sharing the house with others", "I did not feel safe in my building or neighborhood due to the virus", and "Moved to a more comfortable home."

```
. tab home_move_why_3_w1b urban_pre [aw=weight_w1b] if
home_move_why_3_w1b!="Seen but unanswered", col nofreq
```

| Moving reason: shared home | Moved from neighborhood type | | Total |
|---|---|---|---|
|  | Lower-den | Dense urb | |
| I did not feel safe.. | 7.93 | 8.29 | 7.99 |
| Not selected | 92.07 | 91.71 | 92.01 |
| Total | 100.00 | 100.00 | 100.00 |

```
. tab home_move_why_4_w1b urban_pre [aw=weight_w1b] if
home_move_why_4_w1b!="Seen but unanswered", col nofreq
```

|                          | Moved from neighborhood type |           |        |
|--------------------------|-----------------------------:|----------:|-------:|
| Moving reason: virus in neighborhood | Lower-den         | Dense urb | Total  |
| I did not feel safe..    | 9.63                         | 5.39      | 8.96   |
| Not selected             | 90.37                        | 94.61     | 91.04  |
| Total                    | 100.00                       | 100.00    | 100.00 |

```
. tab home_move_why_8_w1b urban_pre [aw=weight_w1b] if
home_move_why_8_w1b!="Seen but unanswered", col nofreq
```

|                          | Moved from neighborhood type |           |        |
|--------------------------|-----------------------------:|----------:|-------:|
| Moving reason: comfortable home | Lower-den             | Dense urb | Total  |
| Moved to a more com..    | 24.12                        | 26.28     | 24.46  |
| Not selected             | 75.88                        | 73.72     | 75.54  |
| Total                    | 100.00                       | 100.00    | 100.00 |